# FlexiSec: A Configurable Link Layer Security Architecture for Wireless Sensor Networks

Devesh Jinwala*[1], Dhiren Patel[2] and Kankar Dasgupta[3]

* corresponding author

[1]Central Computer Centre, Sardar Vallabhbhai National Institute of Technology,
Ichchhanath, Surat, India
dcjinwala@gmail.com

[2]Department of Computer Engineering, Sardar Vallabhbhai National Institute of Technology,
Ichchhanath, Surat, India
dhiren29p@gmail.com

[3]Space Applications Centre, Indian Space Research Organization,
Satellite Road, Ahmedabad, India

**Abstract**: Ensuring communications security in Wireless Sensor Networks (WSNs) indeed is critical; due to the criticality of the resources in the sensor nodes as well as due to their ubiquitous and pervasive deployment, with varying attributes and degrees of security required. The proliferation of the next generation sensor nodes, has not solved this problem, because of the greater emphasis on low-cost deployment.

In addition, the WSNs use data-centric multi-hop communication that in turn, necessitates the security support to be devised at the link layer (increasing the cost of security related operations), instead of being at the application layer, as in general networks. Therefore, an energy-efficient link layer security framework is necessitated. There do exists a number of link layer security architectures that offer some combinations of the security attributes desired by different WSN applications.

However, as we show in this paper, none of them is responsive to the actual security demands of the applications. Therefore, we believe that there is a need for investigating the feasibility of a *configurable* software-based link layer security architecture wherein an application can be compiled flexibly, with respect to its actual security demands.

In this paper, we analyze, propose and experiment with the basic design of such configurable link layer security architecture for WSNs. We also experimentally evaluate various aspects related to our scheme viz. configurable block ciphers, configurable block cipher modes of operations, configurable MAC sizes and configurable replay protection. The architecture proposed is aimed to offer the optimal level of security at the minimal overhead, thus saving the precious resources in the WSNs.

## 1. Introduction

Wireless Sensor Networks (WSNs) comprise of the networked wireless sensor nodes to realize some vital functionality. Some of the example applications of WSNs are *environmental parameter monitoring* (flood, water level, temperature, stress, strain, pressure etc), *industrial automation in hostile environments, tracking movements of living beings in parks, sanctuaries, offices, schools, banks etc, surveillance in war zones, enemy camps* and a host of others [1][2]. Irrespective of the applications for which the WSNs are deployed, the security of the network nodes by themselves and that of the data collected and disseminated by them is of prime concern [2][3].

Wireless sensor nodes are characterized by severe constraints in power, computational resources, memory, and bandwidth and have small physical size with low power consumption [1][2]. On the other hand, using the security protocols always entails additional overhead on the *computational, storage* and *energy* resources. Therefore, ensuring security in WSNs requires *adapting* the conventional security protocols to the resource constrained WSN environments, with minimal overhead.

In addition, it is a fact that *unconditional security* is not possible in practice, even in the resource rich, conventional networks [4]. Therefore, the security protocols therein, have always been designed to offer *computational security* only. Computational security is based on either the *intractability* of solving a particular computational problem or on ensuring that the *computational complexity* of the security algorithm is too high to enable it to be broken, within reasonable time and within the available resources [4]. Therefore, the security protocols in conventional networks are designed to operate and be safe, within the precincts of the available resources on the target platform, only.

Now, in order to have *confidence* in computational security, the conventional security protocols are inherently designed with higher computational, storage and energy resource overheads [5]. Such higher overheads are tolerable by the *resource-rich* Personal Computers, but may not be tolerable by the *resource-starved* sensor nodes. Therefore, in the WSNs, due to the low resources, it is even more essential to ensure that the associated overhead due to the security protocols is at the minimum.

In addition to the above-mentioned facts, the communication paradigm followed in WSNs is *data-centric multi-hop communication*, instead of the *route-centric* *multi-hop* communication used in the conventional networks. The data-centric multi-hop communication is characterized by the *in-network* processing. In-network processing is based on *data fusion* that involves on the fly *aggregation, summarization,* or *duplicate elimination* of the data collected from differentسensor nodes [1][2]. Since the processing of the data is done *on-the-fly*, the overall communication costs are reduced [1]. However, the fall-out of this is that with *multi-hop communication* and the *in-network processing* demanding applications, the conventional *end-to-end* security mechanisms alone are not sufficient for the WSN [3]. This means that the direct applications of the protocols like IPSec [6], TLS [7] SSL [8] in WSNs is obviated.

Therefore, the security protocols in the WSNs have to be devised at the link layer to support *secure* data fusion [3] [9]. In the link layer security protocols, because the security related operations are carried out at every hop, the overhead due to the same further increases [3]. Hence, these protocols have to be designed carefully to reduce the overhead to the extent possible.

In order to investigate whether the overhead can somehow be optimized or not, it is necessary to identify what specific link layer security attributes are desired and whether any refinement is possible with respect to each one of them or not. We discuss in the next section, how in attempt to improve the resource overhead in relation to the security attributes desired, the notion of the *configurable security attributes* is useful.

## 1.1 Motivation

The security attributes that any link layer security protocol must offer are *data confidentiality* (mechanism: data encryption), *data integrity and data-origin authentication* (mechanism: keyed hash functions) and *data freshness with protection against replay attacks* (mechanism: an appropriate anti-replay algorithm).

However, not all the sensor network applications demand all of these attributes. Well, we observed earlier, that the WSNs are deployed for a wide and diverse spectrum of applications. Therefore, there are wide variations in the security demands of various applications, too. Hence, it is essential to identify the classes of WSN applications that demand similar security attributes.

In [10], we elaborately survey a wide cross-section of live WSN applications that have diverse demands for security, using which we propose a *security attributes driven taxonomy* of WSN applications. We use the proposed taxonomy as the basis, to analyze the *level* and the *types* of security attributes desired in each category of applications.

In addition, from the survey of the types of the existing WSN nodes and their configurations, we realize that the available resources in the sensor nodes also vary a lot. For example the Berkeley Mica2 [11] motes, released first in 2004, have 128 KB of program memory and 4 KB of RAM whereas the Crossbow iMote2 motes released in 2007 have 256kB SRAM, 32MB flash memory and 32MB of SDRAM [12].

Hence, we primarily conclude that *the link layer security architecture in WSNs must be responsive to the application environment under consideration*. This means that it should be *flexible* and *configurable*.

Unfortunately, our investigation reveals that the existing attempts at providing *software-based* link layer security support viz. *SPINS* [13], *TinySec* [14], *SenSec* [15], and *MiniSec* [16] assume *abstract security models* of the WSN deployment that are not completely sensitive to the actual security demands of the target applications. Therefore, these frameworks do not allow *configurability* in tuning the *type* and *level* of the security attributes for an application to help minimize the resource overhead.

Alternately, the necessary link layer security support may be provided by the underlying hardware based on IEEE 802.15.4 specification [17][18]. However, using a hardware-based solution lacks flexibility and does not provide the *transparent security enablement* for the existing WSN applications [19].

Therefore, we propose here, model of a *flexible* link layer security architecture viz. *FlexiSec*. We attempt to make the model *configurable* by providing various *security attributes driven alternatives* to the application designer, to enable the optimal target code for the application under consideration, to be generated.

We arrive at the *configuration* of each such *security attributes driven alternative* in the proposed model, by investigating the *type* and the *level* of security attributes, that alternative is aimed to offer. In order to support our design, we implement and analyze the alternative configurations therein and show that our design performs better than the contemporary architectures.

To be specific, our experimentations and analyses are done with respect to the overhead associated with various *block ciphers*, the *block cipher modes of operations*, the *message authentication code sizes,* and the *replay protection* algorithms. In the next section, we present an overview of the specific research issues that helped us in arriving at the configuration of our proposed model; with proper justification that made us focus on those issues.

## 1.2 Scope of Work

The link layer security architecture that we propose here viz. FlexiSec is aimed to offer configurable security with respect to the following aspects.

(a) First, our proposed framework allows application designer tune the desired security attributes necessary for an application. The proposed framework is aimed to support either *message/entity authentication* or a combination of *confidentiality and authentication* or that of *confidentiality, authentication, and replay protection* along with flexible selection of *MAC sizes*. Such flexibility is necessary with respect to the diverse security demands of the WSN applications, as illustrated in [10].

As for example, with the combined Authenticated-Encryption technique that the Output Codebook Mode (OCB) [20], the Counter with Cipher Block Chaining Message Authentication Code (CCM) Mode [21] and the Galois Counter Mode (GCM) [22] follow, the applications demanding the *message confidentiality* as well as *data integrity*, are more efficient to implement, as compared to the same using the conventional modes. This is so, because the conventional modes like Cipher Block Chaining Mode [23] are designed to support confidentiality and not message integrity check. Hence, the message integrity check algorithm like CBC-MAC [24] is required to be employed separately, resulting into increased overall overhead.

On the other hand, if the security architecture employs only the Authenticated Encryption (AE) modes of operation, then the applications demanding only authentication may not be implemented as efficiently, as with the conventional CBC-MAC mechanism.

(b) Next, we propose that in the link layer security framework employed, the *strength* of the cipher used must be commensurate with the *resource overhead* entailed and the *security level* desired. We emphasize that the confidence in a security protocol is largely derived from that in the cipher. At the same time, the total overhead associated with the use of such a protocol is also largely due to the intrinsic design of the cipher and due to the cipher parameters viz. *the key-size, block-size and the number of rounds*.

Therefore, instead of merely using a cipher just because it is accepted as an advanced standard, it is essential to examine whether any other lightweight cipher, operating without any compromise on the necessary cipher parameters (beginning with the key-size), can be feasibly employed or not.

Based on our experiments carried out to measure the overhead associated, we prescribe two ciphers at two extremes; yet offering the security strength due to 128-bit key-size. The prescribed ciphers are the *AES cipher Rijndael* [25] and an optimized version of *lightweight cipher Corrected Block Tiny Encryption Algorithm (XXTEA)* [26]. As part of our experimentation, we propose two optimized versions of the AES viz. a *speed optimized* and a *size optimized* version, too.

We again emphasize that the availability of multiple configurable options to the application programmer, enables him/her to compile an application with the optimal alternatives depending upon the availability of the resources and the security demands of the application; thereby tuning the optimum level of security, at the minimal overhead.

(c) Next, again with an aim to reduce the resource overhead, we propose that the size of the Message Authentication Code (MAC) employed must be commensurate with the packet transmission rate of the application under consideration. The current link layer security architectures [14][15][16] employ 4-byte MAC. We consider varying MAC sizes of 4 bytes or 8 bytes, based on the security attributes driven taxonomy of WSN applications [10] as before.

(d) Lastly, we show that a simple *Bloom-filter* based anti-replay technique for supporting link layer replay protection as we proposed in [27] is useful for FlexiSec; as compared to the conventional methods for implementing replay protection [28].

Hence, the innovative approach in our solution will offer the following benefits as compared to the peer link layer security frameworks for WSNs:
- Our proposed framework has configurable and flexible security features that can be tailored to the needs of the application under consideration.
- It can be used as a ready-to-use experimental test-bed for security related experimentations in WSNs.
- It can offer seamless migration of legacy applications to make them security enabled such that there is no need for separate API calls to be made to use the security features.

As has been the test-bed for our experimentations, the proposed security architecture is also intended to be integrated with the TinyOS [29] environment. However, it can be extended for any other alternate platform.

To the best of our knowledge, ours is the first attempt at proposing a model of *configurable* link layer security architecture that employs and evaluates the XXTEA and the AES ciphers in the CBC, OCB and CCM block cipher modes as well as proposes a simple Bloom filter based replay protection algorithm for unicast communications [28][27] .

The rest of the paper is organized as follows: in section 2 we survey the existing software based link layer security architectures. In section 3, we describe the characteristics of the ciphers and the modes of operations used by us for evaluation. In section 4, we present a survey of the conventional approaches for ensuring replay protection in WSNs and the Bloom filter based approach for replay protection along with our design. In section 5, we discuss our methodology of evaluation and the experimental setup. In section 6, we present the results and analyze them whereas in section 7, we discuss the proposed model of the FlexiSec architecture. We conclude with the plan for the further work in section 8.

## 2. Link Layer Security Architectures

In this section, we present a survey of the existing link layer security architectures. There indeed have been a significant number of attempts published in the literature, aimed at enriching the link layer protocols in WSNs, with the essential security mechanisms.

Karlof et al present TinySec [14] as a lightweight and efficient link-layer security protocol for WSNs. It is the first link layer security architecture implemented in software. TinySec attempts to provide at least minimal configurable link layer security, with three different modes of operations. These operation mode provide either (a) no security support or (b) support for message authentication only based on Cipher Block Chaining Message Authentication Code (CBC-MAC) [24] viz. *TinySec_Auth* or (c) support for (b) and message confidentiality via encryption in Cipher Block Chaining (CBC) mode [23] viz. *TinySec_AE* mode.

The designers of TinySec emphasize that protection against replay attacks need not be built-in at the link layer. Therefore, there is no mode of operation that offers such selection.

Architecturally, TinySec also aims to employ a *plug-and-play* design to make it possible to alter the basic building blocks of the security architecture i.e. the *cipher* employed or the *cipher mode of operation* used. However, we experienced that factually, such alteration requires significant changes to the underlying TinySec code. As for example though it is easy to use a 128-bit key-sized cipher by merely changing a few lines in the TinySec configuration file viz. *TinySecC.nc*, employing a 128-bit block sized cipher requires a careful understanding of the TinySec architecture and substantial cryptic changes to various other implementation modules.

TinySec employs 64-bit block-sized and 80-bit key-sized Skipjack [30], as the default block cipher. The authors of TinySec also discuss the experimental evaluation of the encumbered 128-bit key-sized and 64-bit RC5 block cipher [31].

SenSec is another attempt at designing the link layer security framework for *specific* body monitoring application at the Institute of Infocomm Research, Singapore. Thus, it cannot be employed as a platform for further cryptographic experimentations. It also uses Skipjack as the block cipher but with modified XCBC [32] mode of operation. Being modeled highly on the design of TinySec, SenSec too does not support replay protection. TinySec and SenSec are devised for security support in only unicast communication mode.

Mark Luk *et al* in [16] present a recent attempt at designing the link layer security architecture that is designed for the Telos motes [33]. MiniSec uses a different approach in that it offers two operating modes, one tailored for *single-source unicast* communication, whereas the other, for *multi-source broadcast* communication. It offers all the basic desired link layer security properties viz. data encryption, message integrity and replay protection. MiniSec also uses Skipjack as the block cipher.

However, MiniSec offers no configurability with respect to the security attributes, offering only AE block cipher mode of operation. Thus, applications demanding only message or entity authentication cannot be optimally implemented in MiniSec. Neither does MiniSec offer any choice in selection of cipher as well as the selection of the MAC sizes.

Alternatively, the link layer security support may be provided using a protocol confirming to the IEEE 802.15.4 specification [17]. One of the protocols confirming to the IEEE 802.15.4 standard, is the ZigBee protocol [34]. ZigBee is a specification, targeted at RF applications that require a low data rate, long battery life, and secure networking. However, the use of ZigBee protocol involves appropriate licensing and membership of the ZigBee Consortium. In addition, ZigBee has been labeled as a protocol offering *high security at high overhead* in [16] and hence suitable for the *higher end* PAN devices and not the *low end* WSN nodes. This is factually reflected in its cipher specification – all of the modes in IEEE 802.15.4 specification are based on the AES Rijndael cipher that indeed - with all improved and optimized versions – is still an *expensive* cipher. In addition, the IEEE 802.15.4 specification offers a mode that offers only data encryption using Counter mode [23]. With any encryption of the data without message authentication labeled as being useless, the limitations of such a mode of operation have well been discussed in [35].

As compared to these solutions, the design that we propose here is better due to the following reasons:

(a) In FlexiSec model, we make a provision for three different optimized versions of two different ciphers. The ciphers we have selected are the AES and the XXTEA cipher. Our results indeed show that the XXTEA cipher can be a better choice for those environments where the available storage is limited, whereas the optimized versions of the AES cipher are better otherwise.

(b) We also provide configurable selection of either the conventional CBC-MAC for ensuring authentication only or the efficient OCB Authenticated Encryption block cipher mode of operation for ensuring confidentiality as well as integrity in FlexiSec.

(c) FlexiSec also caters to the selection of variable MAC sizes that are responsive to the application needs and hardware used.

(d) FlexiSec also has the provision of a simple and efficient replay protection schemes.

(e) FlexiSec explicitly omits the insecure block cipher modes of operation that are specified in IEEE 802.15.4 standard viz. the Counter mode without authentication support.

(f) In FlexiSec, we also make the provision of an option for *non-keyed* authentication support as well as the provision of security support being derived from hardware too.

We discuss the detailed model of FlexiSec in section 7.

## 3. Block Ciphers and Modes for FlexiSec

We now discuss the characteristics of the block ciphers and the cipher modes of operation that we believe can be used as potential candidates in the link layer security architecture FlexiSec. We observe that in our process of selecting and/or optimizing the selected cipher and the cipher mode of operation for the link layer security architectures, it is necessary to look for any peer attempts to do so.

This is so because instead of reinventing the wheel, we could then use the same results for our architecture. Hence, in section 3.3, we discuss the related work attempted at evaluation of the block ciphers and the modes in the WSN environment and justify how our approach here, is different from the contemporary ones.

### 3.1 The Block Ciphers

We attempt the evaluation of the AES cipher Rijndael, the XXTEA cipher, RC6 [36] and the Skipjack cipher. Our aim is to identify different possible candidate ciphers to comply with the nature of the application under consideration and the available resources. Hence, we compared the performance of the AES, RC6 and XXTEA against that of Skipjack and tried to justify as to how amongst the ciphers with the key-size of 128-bits, AES and XXTEA can be used as two suitable candidates. As mentioned earlier, we also optimize the AES and the XXTEA ciphers for the link layer security architecture for the WSNs.

Our baseline cipher for evaluation is Skipjack, since it is the cipher of choice, in all the existing software based link layer security architectures. Skipjack uses 80-bit key with a 64-bit block size and 32 rounds of an unbalanced Feistel network. The cipher was declassified in 1998 with an aim to replace then standard cipher viz. the DES [37]. The best cryptanalytic attack against the cipher was carried out on its 31 of the 32 rounds, employing differential cryptanalysis [38].

However, we attempted exploring whether 80-bit key-size of Skipjack can be considered sufficient today? Although the strength of a cipher is a complex measure of the *cipher function, key-size and the block-size*, we believe that the size of the cipher key is primarily an indicative measure of the strength of the *computational security* of the cipher. The cipher key-size must be sufficient enough to prevent the brute force attack against the cipher. Hence, we attempted to analyze the justification for the 80-bit key-size, as we discuss below.

In [39], Lenstra attempts to quantify the security of a cryptosystem with respect to its key length in terms of the trust in the Data Encryption Standard [38] cipher. DES was first introduced in 1977 and then was reviewed every five years. Thus, it was trusted at least till 1982. Lenstra proposes that the security of DES in 1982 be treated as the base security to calculate the *security margin y* of any other cipher. Assuming $k$ as the key-size required to carry out the best known attack against a cipher $A$, the number of years till the cipher $A$ can be considered secure i.e. the security margin $y$ of the cipher $A$ can be defined as

$$y = 1982 + (k - 56)\frac{30}{23} \qquad (1)$$

IF $y$ is known i.e. the year up to which the cipher is required to be secure, the minimum required $k$ can be found out. If calculated in this manner, the 80-bit key-sized cipher can be considered as safe until 2013 and a 128-bit key-size cipher, until 2076.

However, this hypothesis was given by Lenstra in 2001. The recent recommendations portray a very different picture. As per the claims of RSA Security Labs, 80-bit keys would become *crackable* by 2010 [40]. In addition, we found a conservative prescription by ECRYPT [41] that dictates a minimal of 128-bits keys be used, for any cipher.

In the Table-I below we show the prescribed key-sizes by ECRYPT, NIST [42], NSA[43], and RFC3766 [44].

Table 1    Cipher Key-sizes

| Organization /Method | Desirable SKC key-size in bits | Desirable PKC key-size in bits | Secure till year |
|---|---|---|---|
| ECRYPT | 128 | 3248 | 2029 |
| NIST | 128 | 3072 | 2030 |
| NSA | 128 | - | - |
| RFC 3766 | 128 | 3253 | - |

From this discussion, we conclude that it is necessary to look for a 128-bit key-sized cipher even for the low-resource WSN environments. We have considered three categories of candidate ciphers with 128-bit key-size.

First, we have selected AES because it is the current cipher standard. The AES cipher follows the substitution-permutation network structure. Next, we selected RC6 because it was one of the candidate cipher for the Advanced Encryption Standard selection and has been used as a cipher suitable for embedded systems in [45]. On the other hand, in the lookout for 128-bit key-sized lightweight ciphers, we tried selecting a cipher from the Tiny Encryption Algorithm family.

The TEA (Tiny Encryption Algorithm) [46] cipher is a 64-bit block cipher with 128-bit key-size and 64 rounds of operation. It is a short Feistel iteration cipher with no preset tables, nor any explicit key mixing routines. The 64 rounds of Feistel operations are based on the expressions viz.

$$sum = sum + \lfloor(\sqrt{5}-1)2^{31}\rfloor$$
$$RH = RH + (LH << 4 + key[0]) \oplus (LH + sum) \oplus ((LH << 5) + key[1])$$
$$LH = LH + (RH << 4 + key[2]) \oplus (RH + sum) \oplus ((RH >> 5) + key[3]) \qquad (2)$$

The key mixing operation in TEA uses addition and a golden number delta given by

$$\lfloor(\sqrt{5}-1)2^{31}\rfloor \qquad (3)$$

whereas the second and the third expressions show the Feistel functions of the cipher.

Various cryptanalytic attacks on TEA have been reported in the literature. Biryukov and Wagner exposed the slide attack on TEA in [47], whereas Kesley in [48] exposed the reduced key strength of the cipher due to the equivalent keys.

Hernandez et al. in [49] showed that, it is possible to find the distinguisher for TEA with only $2^{115}$ chosen plaintexts and five rounds, using genetic algorithms. Moon et al in [50] show that with $2^{52.5}$ chosen plain texts and $2^{84}$ encryptions, it is possible to attack an 11-round TEA with impossible differentials. In [51], Seokhie Hong use the truncated differentials of probability 1 to attack a 17-round TEA with 1920 chosen plain texts that improves on the previous attacks.

Needham et al. in [52] propose the XTEA cipher, to overcome the limitations of TEA, specifically the related key attack. However, XTEA is shown to be more vulnerable to differential and truncated differential attacks in [51].

Wheeler et al. then propose in [26] the Corrected Block TEA cipher (XXTEA) as an improvement over Block TEA. XXTEA also uses 128-bit key-sizes with a Feistel network design and variable 6 to 32 rounds depending on the block size.

No major published cryptanalytic weaknesses of XXTEA cipher are known. Therefore, we deduce that TEA and XTEA should not be adjudged as the lightweight cipher of choice for our framework. Hence, we selected the XXTEA cipher for our framework. We believe that because of its simplistic design, it ought to be appropriate for the resource constrained WSNs.

## 3.2 Block Cipher Modes of Operation

We have also attempted the optimization and evaluation of the block cipher modes of operations. Commonly the block cipher modes of operation can be categorized as either *Conventional* mode or the *Authenticated Encryption* mode. The Cipher Block Chaining mode is the commonly used conventional mode of operation. The CBC mode encryption with the ciphertext block *C*, the plaintext block *P*, the cipher key *K*, the encryption function *E* and the initialization vector *IV*, is defined as [23]

$$C_i = E_K(P_i \oplus C_{i-1}), \quad C_0 = IV \quad (4)$$

whereas, the CBC mode decryption operation is defined as

$$P_i = D_K(C_i) \oplus C_{i-1}, \quad C_0 = IV \quad (5)$$

The CBC mode necessitates that for every encryption round, the initialization vector must be unique in order to ensure *semantic security*. In CBC, for an input message of *N* blocks, *N* block cipher calls are required only for ensuring confidentiality i.e. there is no computation of the MAC during the encryption round. The MAC can be computed using any other algorithm but normally, it is computed using the CBC-MAC [24] that is defined as,

$$C_i = E_K(P_i \oplus C_{i-1}), \quad C_0 = 0^n \quad (6)$$

Thus, in computation of MAC also, another *N* block cipher calls are required.

In addition, since this is a separate call to the block cipher; different set of keys than the one used for computing the ciphertext (for confidentiality) are required here. Thus, the total overhead not only increases because of the multiple calls to the block cipher but also due to the additional key management, required.

As compared to the same, the Output Codebook Mode (OCB) [20], the Counter with CBC-MAC (CCM) [21] Mode and the Galois Counter Mode (GCM) [22] i.e. the Authenticated Encryption (AE) modes were proposed to lower the computational and storage costs.

Phillip Rogaway et al. proposed the OCB mode in [20]. In this mode, the plaintext is encrypted to get the ciphertext and then the MAC (computed with a MAC computation routine) is appended at the end of the encrypted message in the form of a *tag*. The length of the tag controls the level of authentication.

Formally, first in this mode, an offset or a random value is computed as

$$O = E_K(0^b) \quad (7)$$

where *O* denotes the b-bit zero vector encrypted under the key *K*. OCB mode employs a nonce instead of an *IV* or a *counter* to randomize the encryption. If *N* denotes the nonce for a message

$$M = M_1 M_2 M_3 M_4 \ldots M_n,$$

then the i$^{th}$ block $M_i$ in OCB is encrypted as

$$E_K(M_i \oplus O_i \oplus N) \oplus O_i \oplus N \quad (8)$$

whereas it computes the tag as

$$E_K(M_i \oplus \ldots \oplus M_n \oplus Y_n \oplus O_{n+1} \oplus N) \quad (9)$$

The OCB protocol requires two state variables. First, it employs a single cipher key *K* used for both encapsulation and decapsulation. In addition, it uses a 28-bit packet sequence counter to construct the OCB mode nonce.

For a message of *N* blocks, OCB requires 2 + *N* block cipher calls to encrypt the message as well as compute the tag, in a single invocation of the cryptosystem. Thus, the principal characteristics of OCB are in achieving optimal number of block cipher calls, parallelizable routines and the use of a single-key [20].

Russ Housley, Doug Whiting and Niels Ferguson proposed the CCM mode in [21]. It is based on employing the cipher in the *Counter* mode of operation and then using the CBC-MAC for computing the MAC. The CCM mode is less efficient than the OCB mode as it requires twice the number of block cipher calls, as required by the OCB mode.

John Viega and David A. McGrew proposed the Galois Counter Mode as an improvement to Carter-Wegman Counter CWC mode (CWC) [53]. It is now recommended by the NIST as its latest standard [54].

The GCM mode offers the triple benefits of *being parallelizable in hardware and software*, *being able to authenticate the data beyond those, which are not encrypted* (Authenticated Encryption with Associated Data (AEAD)) as well as *being patent free*. It is specified to be used with 128-bit ciphers only in general and with the AES cipher in particular, although the block size of the underlying cipher should not matter. We have implemented GCM not only with the AES versions but also with the 80-bit Skipjack cipher.

The GCM mode dispels the CCM approach of combining an already existing *encryption mode* with an already existing *message integrity check mechanism* (result: two passes) albeit using a single key to offer authenticated encryption. It even does not use the inverse cipher function of the underlying cipher. GCM uses a conventional mode for encryption but uses the authentication by computing the MAC in the Galois field. It takes as input a block cipher key *K*, an initialization vector *IV*, a plaintext *P*, and any additional data *A* to be authenticated. In turn, it gives as output the ciphertext *C*, and an authentication tag *T*

Albeit for the gain in performance that one obtains using the AE modes, it is emphasized that they are useful only when an application demands confidentiality as well as message integrity both. For those applications requiring only the message authentication and integrity, the AE modes are not useful.

## 3.3 Related Work in Evaluation of Ciphers and Modes

In general, the block ciphers used for evaluation in WSN environment are viz. RC5 [31], Skipjack [30], Rijndael [25], Twofish [55], KASUMI [56], Camellia [56], TEA [46].

There have been many benchmarks and evaluation of the block ciphers for the WSNs as surveyed and discussed here.

Law *et al* in [57], present a detailed evaluation of the block ciphers viz. Skipjack, RC5, RC6, MISTY1 [58], Rijndael, Twofish, KASUMI, and Camellia. The evaluation is based on security properties, storage and energy efficiency of the ciphers. The results prescribe Skipjack as a suitable cipher in low memory resources environment to offer low security strength, MISTY1 as a suitable cipher for higher security strength and AES as a suitable cipher for the applications that necessitate highest speed of operation at higher demands for memory.

However, their evaluation of the ciphers is not done within any link layer architecture, thus not accounting for any link layer framework overhead. In addition, no attempt has been made to optimize the cipher code – instead, simply the openSSL [59] versions of the ciphers are employed. Moreover, as against the recommendation of these results, RC5 has been reported to be having higher speed than AES in [60].

In [61], Großshädl Johann *et al* attempt at energy evaluation of the software implementations of the block ciphers. The authors have considered the ciphers RC6 [36], Rijndael, Serpent [62], Twofish and XTEA. They have used the simulation for the StrongARM SA-1100 processor used principally in embedded systems like cell phones and PDAs.

However, this evaluation is also not done within any link layer security framework; nor the actual deployment on the sensor nodes or any typical WSN platform has been done.

In [63] Guimarães Germano *et al*. discuss another attempt at evaluating the security mechanisms in WSNs. They evaluate the ciphers viz. TEA, Skipjack and RC5 on the TinySec platform with the Mica2 motes. However, in this evaluation, the authors do not prescribe any specific cipher as a suitable cipher; nor do they consider vital ciphers like the AES and the XXTEA.

In [64], Luo Xiaohua *et al*. evaluate the performance of ciphers viz. SEAL [65], RC4 [66], RC5, TEA by implementing the ciphers on the Mica2 motes. The evaluation claims that RC5 is not suited for the WSNs. However, as we do here, neither the AES optimization, nor the secure XXTEA (instead of TEA) implementation & optimization, have been attempted.

In [67], Ganesan Prasanth *et al*. analyze and model the encryption overhead by estimating the execution time and memory occupancy for the encryption and message digest algorithms viz. RC4, IDEA[68], RC5, MD5[69], and SHA-1[70] on various hardware platforms viz. Atmega 103, Atmega 128, Mitsubishi M16C/10, Intel StrongARM SA-110, Intel XScale PXA250 and SUN UltraSPARC II processors. Thus, the focus in this research exercise is to evaluate the performance of various embedded hardware platforms using the ciphers as the basis, rather than evaluating the ciphers by themselves. Even otherwise, the vital ciphers like the AES, XXTEA, and Skipjack are not at all considered for evaluation.

Hence, from the above we conclude that none of these research exercises focuses specifically on the block ciphers and block cipher modes in the link layer framework that we attempt here. These evaluations are not aimed at imparting flexibility to the link layer security architecture. The fallout of this fact is that none of the attempts considers the lightweight ciphers of the TEA family as well as the AES cipher for evaluations, nor is there any attempt at optimizing the implementations of these ciphers, to specifically prescribe the ciphers suitable in different environments.

## 4. Replay Attacks and Anti-Replay Schemes

As mentioned earlier, we believe that one of the vital attributes of a link layer security framework is the protection against replay attacks. Therefore, we analyzed, designed, implemented and investigated the conventional replay protection algorithms in WSN setup and came up with a simple yet efficient Bloom filter [28] based anti-replay algorithm as discussed in [27]. In this section, we analyze our replay protection algorithms, intended to be used in FlexiSec.

### 4.1 Replay Attacks

A replay attack occurs when an adversary captures a packet in one protocol run and replays it later, in some other protocol run. Formally, let a sender $A$, at time instant $t_1$ has sent $k$ packets in the set $T$, bearing sequence numbers viz.

$$T_{SEQ} = \{SEQ_1, SEQ_2, SEQ_3, ......SEQ_k\}$$

An adversary orchestrates the replay attack, by capturing any of these packets during the transmission and replaying at an instant later than $A$.

The protection against replay attacks can be devised by defining a *temporal* or *causal* relationship, between the properties of the message received at an instant and those of the messages received previously. Such relationship is exploited to define an anti-replay protocol to handle a replay attack.

Various approaches based on the above are analyzed in [71], [72], [73], and [74]. We survey these with a detailed analysis in [27].

For FlexiSec, we design, implement and evaluate four different algorithms for replay protection viz. based on employing a Counter and a Hash function as state identifiers, and a third based on employing a Bloom filter for detecting replays. We discuss the design, and implementation of the same in [27]. Here, with the focus on FlexiSec design, we show the results obtained to justify the selection of Bloom filter based algorithm for FlexiSec.

*4.1.2    Replay protection in WSNs*

As discussed earlier, the existing attempts at devising link layer security framework in the WSNs are the SPINS [13], TinySec [14], SenSec [15], and MiniSec [16].

SNEP as part of SPINS achieves replay protection by keeping a consistent counter between the sender and receiver. However, as we show in our evaluation, a counter based approach is grossly inefficient in WSNs.

The designers of TinySec do not consider replay protection as desirable at the link layer. They affirm that the replay protection ought to be the responsibility of the application layer.

However, as we analyze in [27], with application layer replay protection, a replayed packet injected into a WSN may travel through many hops before being detected at the destination. This would waste the precious resources in sensor nodes. Hence, it is essential to provide replay protection at the link layer itself.

MiniSec is a secure link layer protocol that serves to provide, according to its designers, low energy consumption and high security. MiniSec provides two different strategies for anti-replay for its two different operating modes.

For the unicast mode, a *synchronized counter* based approach is proposed. However, such scheme is not scalable with the increase in the number of nodes in the network. In addition, it requires costly *resynchronization* routines to be executed, when the counters shared, are desynchronized due to the out-of-order delivery of the packets. As against that, our sliding window based counter approach does not require any resynchronization to be done although it is not scalable, as in MiniSec.

To handle scalability concerns, MiniSec indeed employs a Bloom Filter based approach with sliding window (in broadcast setup). However, our solution against replays is a simpler one, being applicable in a unicast setup in WSNs.

### 4.2 Our Design

In this section, we illustrate and analyze how our Bloom filter based algorithm is efficient and scalable as compared to the conventional approaches.

As mentioned earlier, using a counter value as the message tag is one of the most common approaches for detecting packet replays. However, when we use a counter-based algorithm, we have to devise a suitable data structure to maintain the counter value and incorporate it as one of the data packet component.

| DEST (2) | AM (1) | LEN (1) | SRC (2) | CTR (2) | Data (0..29) | MAC (4) |
|---|---|---|---|---|---|---|

**Figure 1.** TinySec_AE Packet Format [14]

In the TinySec packet format as shown in Fig. 1, a *counter* (*CTR*) indeed is used by its designers for framing the Initialization Vector. Therefore, TinySec can easily be augmented with replay protection using the *CTR* for the purpose. The counter-based algorithm for replay protection is shown in Fig. 2.

```
Algorithm
CounterReplayDetect(CounterReceived,
                                    NodeID)
{
1.  id = o;
2.  for id = 1 to lastValidId {
3.    if (id==NodeID) {
4.    if (CounterReceived <=
                      LastCount[NodeID])
5.      replayed = 1;
        else
6.      replayed = 0;
7.    LastCount[NodeID]=CounterReceived;
      }
   }
}
```

**Figure 2.** Antireplay algorithm using a CTR [IJCS]

However, in a typical WSN, each sensor node will have to maintain the counter value for all the other *(n-1)* nodes in the network. Thereby, the total storage expended turns out to be

```
n(n-1) * bytes_per_counter_value     (10)
```

Therefore, this approach is not scalable and can be employed only for small size networks.

However, this approach does not require counter resynchronization as in SNEP or in MiniSec unicast approach. The counter value required for replay protection is part of IV, so there is no extra overhead in packet size involved for this method. Hence, the method incurs very less consumption in radio energy.

Unlike the previous approach, in the hash-function based algorithm, a hash value generated by a hash function, can be used as a type tag to maintain the state information required for detecting replays. A standard hash function like SHA-1 [70] can be used for the purpose. However, SHA-1 hash function and hence this algorithm entails lots of resource overhead, as we illustrate through our experimental results.

As compared, the Bloom filter [28] based algorithm, which uses a bloom filter with hash functions, is scalable as well as efficient.

The Bloom filter is a special compact data structure used for probabilistic representation of a set so as to probabilistically answer membership queries about the elements of the set. The Bloom filter uses a *Bloom filter vector* as the representative of the data set. The contents of the *Bloom filter vector* are determined by the application of the hash functions on the input data element. That is, the hash values generated upon the application of hash functions to a data item, are treated as different addresses in the Bloom filter vector, the contents of which are set to a 1, to designate the presence of the corresponding data item.

Formally, a Bloom filter is a vector of *n* bits consisting of each individually addressable cells viz. $a_1, a_2, a_3, a_4.....a_n$ along with *m* different hash functions viz. $h_1, h_2, h_3,..........h_m$. Initially, the Bloom filter is empty with all its bits set to 0.

To add the information about a data element, say $d_i$, to the Bloom filter, each of the hash functions $h_j$, $1<=j<=m$ is applied on $d_i$ to get the corresponding hash values $hv_j$. The vector addresses indexed by $hv_j$ in the Bloom filter vector are then set to 1.

Subsequently, to query for the membership of a data element $d_i$, the data element is again hashed with *m* hash functions to get the *m* different hash values viz. $hv_1, hv_2, hv_3,..........hv_m$. Whether an element is a member of the set or not is determined as per the algorithm in Fig. 3.

In using the above approach for replay detection, when a packet is first received by a node, the *message freshness identifier (tag)* in the packet is hashed by the *m* hash functions as before, and the bits in the Bloom vector are set. If these bit values are 1, then the packet is a replayed packet, otherwise it is not, as illustrated in the algorithm in Fig. 3.

```
Algorithm MembershipTests(DataItem a_i,
    BloomFilter vector, Set HashFunctions)
{
1.  for HF = 1 to m {
2.      if (vector[HashFunctions_HF(a)] != 1)
3.      return NonReplay;
4.  }
5.  return Replay;
}
```

**Figure 3.** Generic Algorithm using Bloom filter

### 4.3 False Positives in Bloom filter

Bloom filter based replay protection algorithm may suffer from false positives. That is, even if a packet is not a replayed packet, it may be tagged as a replayed one.

Let the size of Bloom filter vector be $m$ and there be $k$ hash functions, in use. Then, the probability that any one of the location in the vector not being set by any one hash function is given by P'

$$P' = 1 - \frac{1}{m} \tag{11}$$

When employing $k$ hash functions, the same will be

$$P' = \left(1 - \left(\frac{1}{m}\right)\right)^k \tag{12}$$

If there were in all $n$ elements, then the probability for a certain bit to be still not set would be given by

$$P' = \left(1 - \left(\frac{1}{m}\right)\right)^{kn} \tag{13}$$

Then, the probability $P$ that the bit is set will be given by

$$P = 1 - \left(1 - \frac{1}{m}\right)^{kn} \tag{14}$$

Thus, when carrying out the set membership tests, for the data item that is already in the set, the expression (14) above can be used. However, for a data item that is not in the set, the set membership tests must return test above as 0. If we assume that, the set membership test, erroneously returns the same element to be actually present in the set, it would be so only if each of the $k$ hash functions, yield the same value to be 1. That is the probability for a false positive can be derived from expression (14) as

$$P_{False-Positive} = \left(1 - \left(1 - \frac{1}{m}\right)^{kn}\right)^k \tag{15}$$

$$P_{False-Positive} = \left(1 - e^{\frac{-kn}{m}}\right)^k \tag{16}$$

$$\approx \frac{1}{2^k} \tag{17}$$

In general, to prevent false positives, the Bloom filter design criterion specifies tuning $k$ i.e. the number of hash functions.

Our Bloom filter based algorithm for replay protection employs eight hash functions. We show our improved version of the algorithm (as compared to the one in [27]) using eight hash functions in Fig. 4.

In this algorithm, we use bloom filter with eight 32-bit hash functions i.e. theoretically $2^{32}$ different values per hash function. Thus, the false positive rate in our approach turns out to be 1/256 i.e. 0.00390. The eight hash functions that we have employed are from the family of universal hash functions, requiring lesser computational overhead than SHA-1 viz. *RSHash, JSHash, PJWHash, ELPHash, BKDRHash, SDBMHash, DJBHash, DEKHash*, and *APHash*. We carefully implement the bloom filter as two-dimensional vector in order to implement it, optimally.

## 5. Experimental Setup

In this section we describe the tools and the platform used by us for the experiments conducted as well as the methodology adopted and the test application used to do so.

### 5.1 Platforms and Tools used

We have used the TinySec link layer security framework in the TinyOS 1.1x operating environment [29] with the nesC [75] as the language of implementation. We have also employed TOSSIM simulator [76] to first simulate our code and subsequently deploy the code on the Mica2 sensor nodes. We have exploited the TinyOS support for the deployment of the code on the Mica2 motes.

Although TOSSIM captures TinyOS behavior at very low level, it does not model the power consumption for the motes. This is because it does not model the CPU execution time, and thus, cannot provide accurate information for calculating the CPU energy consumption.

Hence, we are using AVRORA [77] to measure the CPU cycles and power consumption for a particular node. Avrora is an emulator implemented in Java that runs the actual Mica code, while emulating each WSN node as its own thread.

Having implemented the ciphers, modes and the replay protection schemes in nesC and plugged them in the TinySec framework, our evaluation in general consists of three different stages, as follows:

```
Algorithm ReplayBloomFilter8(Packet In,Set
    HashFunction,int n, Filter BloomFilter)
{
1. if (!FilterCreated) {
2. Filter CreateBloomFilter (set A, Set
       HashFunctions, integer n, BloomFilter
                                 BloomVector)
3.    FilterCreated=1;
   }
4. for each received packet {
5.    for i = 1 to k {
6.
  HashValue=ApplyHash_i(TypeTag(Packet In))
7.      if (BloomVector[HashValue] == 1)
8.         replayed=1;
9.      else
10.        replayed=0;
11.     BloomVector[HashValue] = 1;
12.     exit();
    }
  }
13. if (replayed == 1)
14.    return (replayedpacket=1)
15. else {
16.    return (replayedpacket=0);
     }
  }
```

**Figure** 4. Antireplay Algorithm `ReplayBloomFilter8`

(a) First, we evaluate the performance in TinyOS environment with TOSSIM as the WSN simulator. The nesC compiler itself gives as output the RAM and ROM requirements of the application under consideration as part of the compilation process that helps us in determining the storage requirement.

(b) Next, we use the Avrora simulator to determine the throughput in bits/sec and the energy consumed. For the purpose, we converted the executable file generated by Mica2 compilation into ELF file format and then simulated it in AVRORA by setting the appropriate flag monitors. We measured throughput according to following formulae with the clock speed of 8 MHz.

$$Throughput = (Message\_size * CPUClockSpeed) / NoofCycles$$
(18)

To measure the energy consumption, we follow the same method as for CPU cycle measurement again using the appropriate flags in Avrora. The total simulation time was set to 100 seconds. We obtain the CPU and radio energy consumption for each mote and normalize the same, for evaluation.

(c) Third, we deploy the application under consideration on the Mica2 motes to verify that the application indeed behaves as examined earlier.

One question which arises here is why did, we employ Mica2 motes, when resource enriched *next generation* motes like Intel iMote [12], Crossbow Iris motes [78] and Moteiv's Tmote Sky motes [79] are available, today. Well, these motes indeed have higher computational and storage power. However, we believe that our evaluation that is carried out on more stringent environment of Mica2 motes, can always be true in more resource-rich environments.

### 5.2  The Test Application

For all of our evaluations we employ a simple application that comes bundled with the TinyOS environment viz. *TestTinySec* – the pseudocode shown in Fig. 5.

```
Algorithm TestTinySec
1. counter = timer
2. while (counter == fired){
3.   if (Send(Data Packet)) then LED=green
4.      else if( Receive(Data Packet))
then
                                  LED=red
  }
```

**Figure 5**. The TestTinySec Algorithm

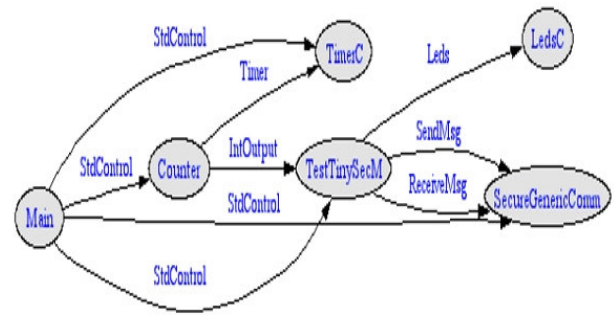

**Figure 6.** TestTinySec application in TinySec

The pseudocode of the application is as shown in the code snippet in Fig. 5 whereas its call-graph generated under the nesC compiler is shown in Fig. 6. *TestTinySecM* is the main module of the application. It uses a counter (which is part of the data packet sent/received over the radio) that is incremented on firing of a timer.

*TestTinySecM* module also uses the TinyOS interfaces *SendMsg* and *ReceiveMsg* those are implemented by the component *SecureGenericComm*. *SecureGenericComm* is responsible for sending and receiving the secure messages over the radio. The counter value modified by the component *Counter*, is further passed by *TestTinySecM* (the suffix M here, as per the nesC conventions signifies the main module of the application) through the *SendMsg* interface of TinyOS, for onward transmission over the radio, to the component *SecureGenericComm* of TinyOS. In addition, when the message is sent, the *Leds* interface is used to toggle the LED on the mote. When the message is transmitted by a mote, the LED is turned green whereas, when the message is received by a mote, the LED is turned red.

The module *TinySecM.nc* handles all the security related operations in TinySec. In Fig. 7, we show the partial call-graph showing the security components of the TinySec that come into play, during the execution. The call-graph was generated for the Skipjack cipher wired in the CBC mode that is the default configuration of TinySec. As can be seen, *TinySecM.nc* uses the modules *CBCModeM.nc* and *CBCMAC.nc* with appropriate calls to the default cipher component *SkipJackM.nc* to encrypt and decrypt the message and to generate and verify the MAC. Thus, *SkipJackM*, *CBCMAC* and *CBCModeM* components are not implemented by us.

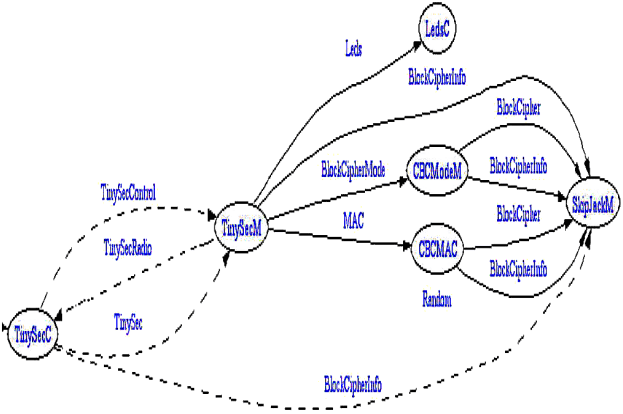

**Figure 7.** Partial Call-graph of the application with Skipjack in CBC Mode

### 5.3 The nesC Versions and Application Call-graphs

As mentioned earlier, as part of its functionality, `TinySecM.nc` eventually calls the default cipher component `SkipjackM.nc`. In our experimentation, we replaced `SkipJackM.nc` with our own appropriate cipher components viz. `TEAM.nc` (for TEA), `XTEAM.nc` (for XTEA), `XXTEAM.nc` (for XXTEA), `OXXTEAM.nc` (for our own optimized XXTEA implementation), `RC6M.nc` (for our RC6 implementation), `AESSPO.nc` (for our own *speed-optimized* AES implementation) and `AESSIO.nc` (for our own *size-optimized* AES implementation).

The 64-bit key for the Skipjack cipher in the default TinySec configuration, is stored in the `./tinyos-keyfile`. On the other hand, the ciphers of the TEA family are all 64-bit ciphers, using 128-bit key-size. Thus, we modified the `./tinyos-keyfile` augmenting the keys in it. In addition, since the block cipher calls for encryption and for computation of the MAC are different calls, two sets of 128-bit keys are required to be stored in the key-file. Execution of these ciphers required modification of the key-size, cipher and mode contexts parameters in the TinySec header file, too. Apart from these files, we also modified the TinySec configuration files viz. `TinySecM.nc`, `CBCMAC.nc` and `CBCModeM.nc`, to lend the 128-bit key support.

We also implement the nesC components for block cipher modes viz. the CCM (`CCMM`) and the OCB mode (`OCBM`) and the GCM (`GCMM`).

We have tried to *size-optimize* the AES version inspired from the C-versions of AES in [80] and have tried to *speed-optimize* the version of AES, inspired from the openSSL [59] version. We subsequently converted these into nesC.

For the AES size-optimized cipher implementation (AESSIO) for our work, we are expending in only two 256-byte wide entries in the F-table and S-table of AES [25]. We implement the rest of the computations required in AES, in the form of in-line functions and macros. Therefore, we expect that the throughput of this approach must be lower as compared to an approach in which the computations are reduced by expending more storage.

In our AES speed-optimized version (AESSPO); we are using the *AES small tables* approach. In this approach, the intermediate values stored in the table are one 32-bit 256 entry forward table and a reverse table each (1 KB + 1 KB) and a reverse S-Box (256 bytes). Thus, the total expended storage in this case is 2048 + 256 = 2304 bytes.

Alternately, when employing *AES large transformation tables,* four forward and reverse tables each with 32-bit, 256 entries are required (4 KB + 4 KB = 8192 B). Thus, the total increased expense in storage is 8192 – 2304 = 5888 B i.e. an increased storage overhead of 71%. For one round, this approach requires 16 table lookups and twelve 32-bit exclusive-OR operations, followed by four 32-bit exclusive-OR operations in the *AddRoundKey* step.

Comparing our size-optimized version with the speed-optimized version, we get a saving of 2304 -512 = 1792 Bytes i.e. a percentage saving of 77.77%. Such saving is indeed significant in the resource-starved environments in the sensor nodes.

The relative comparison of our size and speed optimized versions of AES is shown in Table II. As we analyze in the next section , our theoretical evaluation as done above is actually reflected in the performance evaluation done by us.

Table II  AES Optimizations

| Sr No | Cipher | Storage Expended in Tables | Total Storage |
|---|---|---|---|
| 1. | AES Speed Optimized cipher | 32-bit 256 entries Forward computation table<br>32-bit 256 entries Reverse computation table<br>8-bit 256 entries S-box | 2304 bytes |
| 2. | AES Storage Optimized version | 8-bit 256 entries Forward S-Box<br>8-bit 256 entries Reverse S-Box | 512 bytes |

For XXTEA, we used the basic C-version in [26] and converted it into the nesC version. We also attempted optimizing the basic XXTEA operation using pre-computed tables and inline assembly code. We utilized tables for storing the value of *DELTA* (the magic constant) and round key generation for each round. We used the inline assembly code for the conversions viz. four-byte *character ↔ long* data type. We evaluate both the basic XXTEA cipher (XXTEA) and our own optimized XXTEA (XXTEAO).

We subsequently modified the TinySec configuration files to execute the TestTinySec application using all the combinations of cipher and their modes of operation viz. *AESSpeedOptrimized, AESSizeOptimized* in the OCB, CCM and GCM modes and the Skipjack in the CBC mode.

In Fig. 8, we show the sample partial snapshots of the *TestTinySec* call-graph with XXTEA cipher in the CBC mode, whereas we show the partial call-graph of the application with AES cipher in the OCB mode in Fig. 9.

In addition, since AES and RC6 are 128-bit ciphers as compared to the 64-bit Skipjack and XXTEA, we made appropriate logical changes in the TinySec files, for obtaining this support. For XXTEA, RC6 and AES, we also changed the default *tinyos-keyfile* to enable the support for 128-bit cipher keys.

For the evaluation of our replay protection algorithms also, we use the same test application.

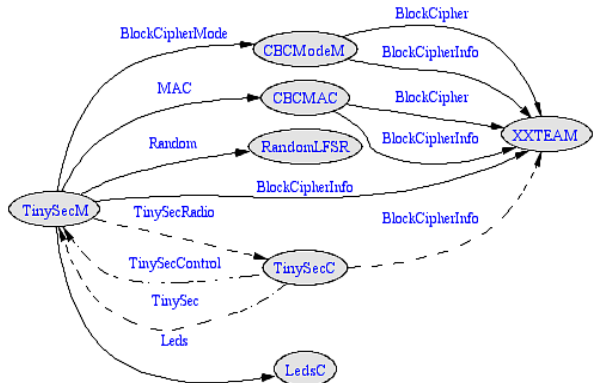

**Figure 8.** Partial Call-graph of the application with XXTEA in CBC Mode

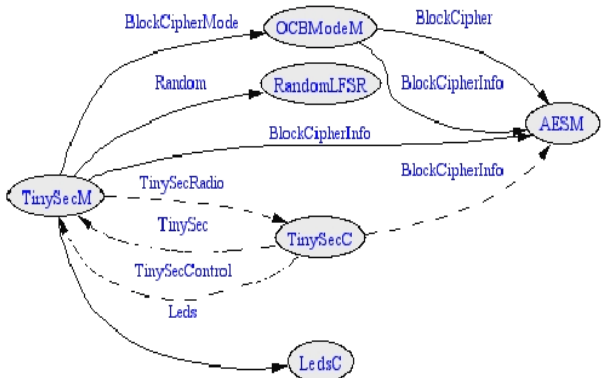

**Figure 9.** Partial Call-graph of the application with AES in OCB Mode

## 6. Results and Analysis

In this section, we first show our experimental evaluation of the block cipher modes of operation, evaluation of ciphers and then the results for the anti-replay algorithm.

### 6.1 Evaluation of the Block Cipher Modes

We present the results of evaluations of the storage requirements (RAM and ROM), throughput in terms of bits/sec and the energy requirements for CBC, OCB, CCM, and GCM block cipher modes in the figures 10 to 13. We must emphasize that since we wanted to examine the advantages and suitability of the block cipher modes of operations, we used only a *single* cipher i.e. AES for evaluation of the cipher modes. The use of AES definitely yields improved security strength as compared to the TinySec default Skipjack cipher. We have used the Skipjack cipher in CBC mode as the reference for our comparison. Hence, it must be emphasized that without looking into any of the metrics, the fundamental advantage that we get using the AE modes is that the latter offer two security attributes as compared to the CBC mode that only offers confidentiality.

From Fig. 10, for the MICA2 motes with only 4KB of RAM, we can see that an overhead of only 13.46% and 12.98% results when using the OCB or CCM mode with the AES speed-optimized (*AESSPO*) and AES size-optimized (*AESSIO*) cipher implementations.

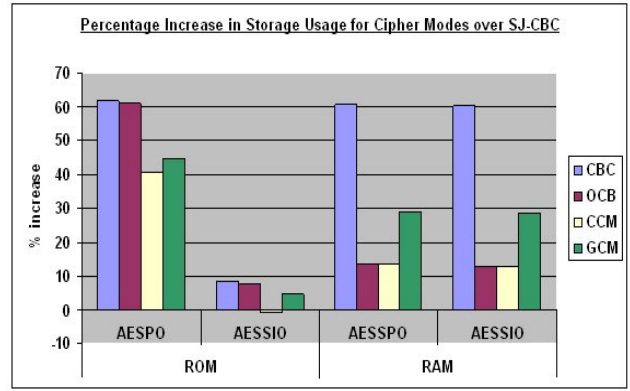

**Figure 10.** Increase (%) in RAM/ROM over the Skipjack in the CBC mode

The same in case of GCM mode is substantially higher. This fact can be a critical factor in mode selection since RAM is scarce even in the recent *next-generation* motes.

However, the significant advantage here is that with under 15% additional overhead in storage, both the attributes viz. confidentiality and authentication are obtained whereas with CBC mode only confidentiality is obtained.

From Fig. 11 for the CPU cycles comparison, we can see that the CBC, OCB, CCM, GCM modes of operation require 129.08%, 48.09% and 133.31% and 54.88% more CPU cycles respectively, when employing AES speed-optimized block cipher over the same with Skipjack cipher in CBC. The corresponding figures for the AES size-optimized version are similar. Thus, when employing OCB mode, the penalty in terms of increased CPU resources is much lesser as compared to the same, when employing CBC/CCM/GCM modes.

In Fig. 12, we show the penalty in the form of lesser throughput when employing the AES cipher. Again, when employing our version of the *AESSpeedOptimized* in OCB, the percentage reduction in throughput is minimal, as compared to the CBC, CCM, or GCM modes.

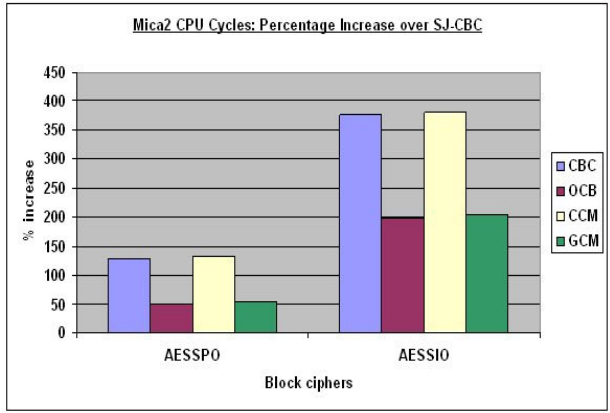

**Figure 11.** Increase (%) in CPU cycles over the Skipjack in CBC mode

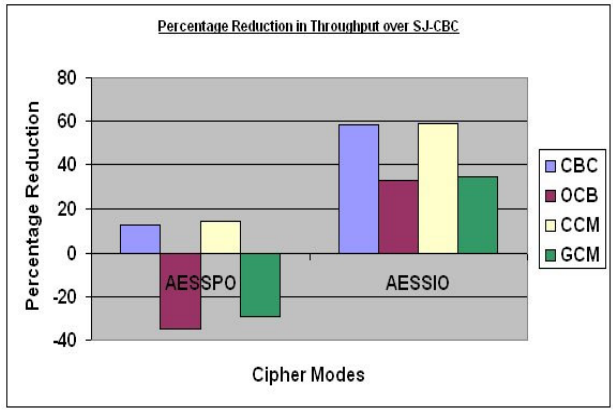

**Figure 12.** (%) Penalty in throughput with OCB/CCM (AES) over CBC (Skipjack)

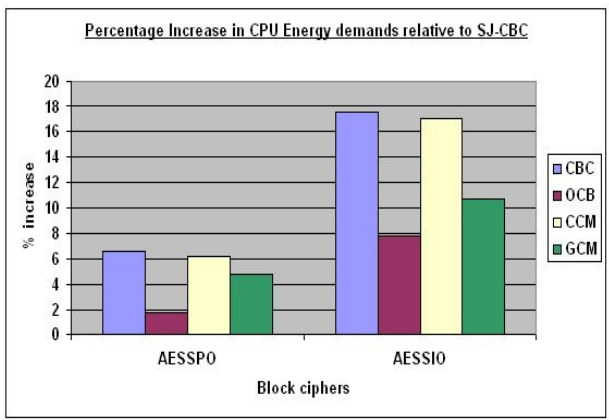

**Figure 13.** (%) Penalty in Energy consumption with OCB/CCM (AES) over CBC (Skipjack)

This is a significant observation pointing to the fact that security due to 128-bit key strength with a stronger mode like OCB is obtained at much lesser overhead. Hence, the OCB mode should be the preferred mode of operation and not the CCM mode as in IEEE 802.15.4 standard specification – at least for the open source link layer security architecture, that we propose here.

Energy is the most critical of all the resources in the WSNs. As can be seen from Fig. 13, the penalty in energy consumption in Mica2 motes for OCB/CCM/GCM modes employed in 128-bit key-size cipher, is much less as compared to the Skipjack cipher in CBC mode.

Even with the increase in the CPU and storage resources in the next generation motes as was pointed out earlier, the energy availability for these motes has almost remained the same as the Mica2 motes. Hence, our results indicating lesser penalty in energy consumption for OCB/CCM/GCM modes are even more significant and vital.

We must again emphasize that the *penalty* when employing the AE modes, referred to above – whether in terms of the increased storage, or increased CPU cycles, reduced throughput, or increased energy consumption, is to be weighed in proper context. For, in return for the same, the underlying cryptosystem will offer two security attributes instead of a single one (when employing only CBC mode) and hence ultimately it results into overall reduction in overhead associated.

### 6.2 Evaluation of the Block Ciphers

In the figures 14-16, we show our results for the cipher evaluation. In Fig. 14, we show the storage requirements for the ciphers under consideration. We observed the RAM requirement of the cipher XXTEA and that of our own optimized XXTEA is the minimal- again which is a significant advantage because the size of RAM has not grown as significantly even in the next generation motes.

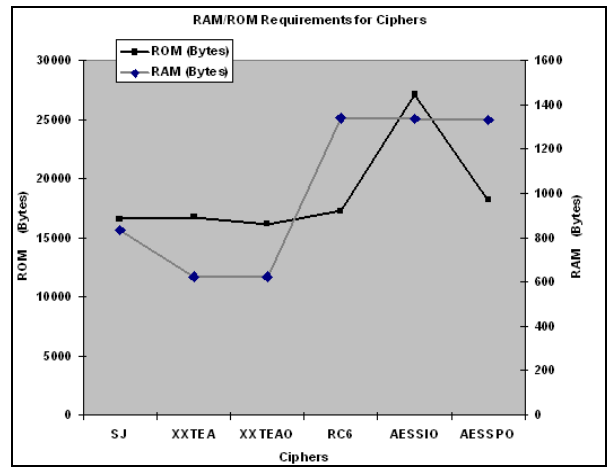

**Figure 14.** RAM & ROM requirements for ciphers

We also show the reduction in the ROM footprint that we were able to achieve in the XXTEAO, with the inline code and the pre-computation of the tables and delta values.

In Fig. 15, we show the energy requirements of the ciphers estimated using the Avrora. From the observed energy requirements of the ciphers, the following points can be concluded:

(a) As compared to Skipjack (80-bit key-size), the energy requirement of the 128-bit key-sized lightweight cipher XXTEAO (our own optimized XXTEA version) is only slightly higher.

(b) The energy demands of AESSPO cipher is quite low, but it requires higher memory.

Since energy constraints in motes are always severer, the XXTEA cipher gives the best of both the worlds – it requires lower memory, provides the security strength due to 128-bit key-size and demands lower energy resources as compared to other 128-bit key-sized ciphers like RC6 and AES.

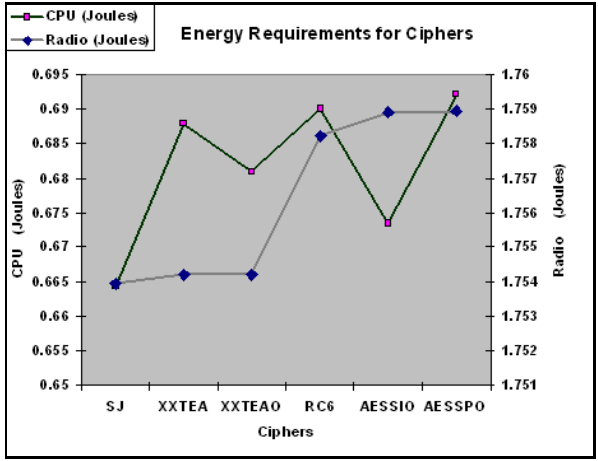

**Figure 15.** Energy in Encryption/Decryption for ciphers

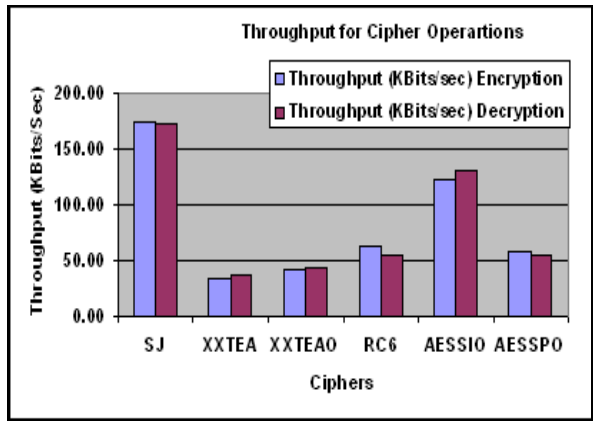

**Figure 16.** Throughput of Cipher Operations

However, this also means that if the available resources permit, the security strength due to the AES standard can also be availed of, by using the speed optimized version of AES.

Nevertheless, as is seen from the observations in Fig. 16, the throughput of XXTEA is the least. We are able to improve this in our own optimized version. However, in general the lesser throughput is due to the large number of rounds (32) involving combinations of rotations and exclusive OR operation.

It may be noted that XXTEA is a cipher not having any *substitution* operation – it achieves the required non-linearity through only the combinations of additions modulo $2^{32}$, rotations and ex-OR operations. While the designers of XXTEA proclaim having considered substitution also as one of the options, it is worthwhile (with increasing storage availability in motes) to investigate a combination of substitution and rotation based operations in XXTEA with lesser number of rounds, than present.

Hence, depending upon the availability of the resources either of the XXTEA or the AES versions can be employed to attain the higher security strength – thus justifying the flexibility in selection of block ciphers.

### 6.3 Evaluation of the Replay Protection Scheme

Again, in order to evaluate the proposed Bloom filter based anti-replay scheme we use the same strategy and the same test application. We have evaluated the replay protection schemes based on different metrics viz. memory, energy and CPU cycles.

We justify the lack of scalability in the Counter-based or Hash-based approaches from the RAM requirements with varying number of neighbors and fixed window-size (of 8). As the results in Fig. 17 show, the graph is almost linear as increasing the number of neighbors, increases the storage requirements, for maintaining the state of previously received counter values. We maintain the state in 2-byte counter in our counter based approach. In SHA-1 based approach we do so in a 20-byte message digest for all the neighbors. Hence it is clear that the memory requirement for SHA-1 based approach is higher than that for the counter based approach. We experienced that the counter based approach works for a maximum of 150 motes, whereas the SHA-1 based approach works for maximum 7 motes with fixed window size of 8. Therefore, these approaches lack the scalability as well as the performance desired.

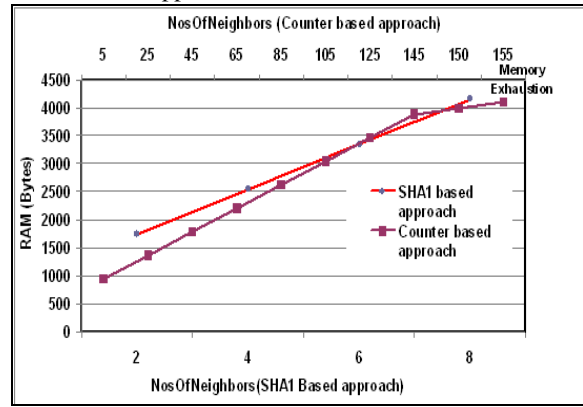

**Figure 17.** RAM requirements for Counter and Hash-based algorithms

In Table III, we give a more meaningful and relative interpretation of the RAM requirements of all of our approaches, using TinySec without replay protection as the basis for comparing the percentage increase in RAM for all the approaches in a network with 10 nodes. It is clear from this table that Bloom filter based approach (with eight hash functions) requires less than 10% overhead in memory, as compared to TinySec without replay protection. Hence, it gives good performance.

**Table III.** Percentage Increase in RAM for all approaches

| Method Description | RAM usage (Bytes) | % increase over TinySec |
|---|---|---|
| TinySec without replay protection | 840 | - |
| Counter Based approach | 1050 | 25 |
| Hash based approach | 4958 | 489 |
| Bloom filter with multiple hash functions | 904 | 7.62 |

The results obtained for the CPU cycle requirement of our implementation, using Avrora are as shown in Fig. 18. Again, the computational overhead for SHA-1 results in higher number of CPU cycles required for all SHA-1 based approaches.

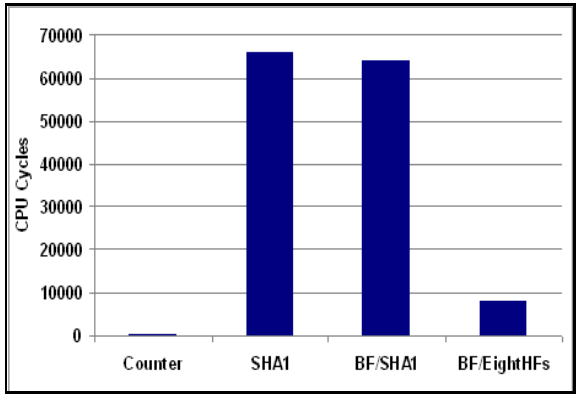

**Figure 18**. The CPU cycles usage for all approaches under evaluation

The results for the energy requirements of our implementation are shown in a summarized form in Table IV, showing the percentage increase in the energy overhead for all approaches over TinySec without replay protection. It is clear from our results that our first and last approaches incur just 0.06% and 0.17% overhead in energy. As mentioned before, the counter based approach lacks scalability. Hence, the Bloom filter based approach offers the best of both worlds – demanding lesser energy and storage, being scalable and at the same time offering replay protection. Along with the lower false positive rate due to eight hash functions (equation 17), the Bloom filter based replay protection algorithm is most suitable.

Table IV. Percentage Increase in Energy Consumption for all approaches

| Method Description | % increase in energy over TinySec |
|---|---|
| TinySec without replay protection | - |
| Counter Based approach | 0.06 % |
| Hash based approach | 4.14 % |
| Bloom filter with single hash function | 3.91 % |
| Bloom filter with multiple hash functions | 0.17 % |

## 7. FlexiSec Architecture

Based on the performance results and the analysis, we arrived at the FlexiSec architecture with the configurable alternatives therein shown in Table V. We propose to modify the TinyOS operating framework with the augmentation of the secure communications stack. Our design is unique in the fact we propose the secure communication stack to be configurable depending upon the data rate, level of security desired and the nature of the application (i.e. the security attributes desired).

As can be observed from the table V, we propose nine different modes of operations. The modes we propose are based on the following observations from our performance results earlier:

Observation 1. We propose that the OCB Authenticated-Encryption block cipher mode of operation, be used for the *security-sensitive* applications in *military, health and socioeconomic domains* e.g. tracking the movement of an enemy troop OR monitoring the health parameters of a patient, OR in human security systems.

In such applications, confidentiality of data is also vital along with authentication. Therefore, the applications by default, demand both these security attributes. When both the authentication as well as encryption operations are required, it is essential to use the AE modes of operations, in order to reduce the nos. of block cipher calls. As our results indicate, the usage of AE modes lends optimal storage, improved CPU utilization and lesser energy consumption.

Since, OCB mode has been found to be the most energy efficient, we propose OCB as the designated AE mode of operation, instead of CCM or the NIST recommended GCM.

However, for the typical *environmental or habitat monitoring applications* like tracking the movement of an animal in a sanctuary OR monitoring the amount of rainfall in the catchments areas of a river across a dam, to enable forecasting the probability of rainfall downstream; the confidentiality of data packets transmitted en route to the base station, is not essential. However, the data integrity, entity authentication, message freshness and replay protection are very vital for the same applications. Hence, the conventional CBC mode can still be employed here.

Observation 2. Our results clearly indicate that the use of the AES standard ciphers entails penalty – either in the form of reduced speed or storage. On the other hand, lightweight cipher XXTEA, offers reasonable security due to its 128-bit key-size. Hence, for all the modes of operations in FlexiSec, we propose that either XXTEA or AES ciphers can be employed – depending upon the available resources and the application.

Observation 3: Lastly, for different applications under consideration, the frequency of packets transmission also would vary. The frequency of transmission of the data packets and the associated radio bandwidth has a very significant bearing on the number of bytes employed for the MAC. Normally, for the WSNs, a 4-byte MAC is employed.

However, it is essential to analyze whether a 4-byte MAC can be considered to be safe from attacks? With a 4-byte MAC an adversary would have 1 chance in $2^{32}$ attempts, to forge the MAC. Now, if the radio is operating at 19.2 kbps, as it is in case of CC1000 based MICA2 motes, one can send only about 40 forgery attempts per second, for the 68-bytes sized TinyOS/TinySec packets. Therefore, in order to make $2^{31}$ tries, the adversary will need at least 621 days to be able to try forging the MAC.

From this discussion, we generalize the model for MAC-forging as explained below. If $k$ is the size of the MAC in bits, $P$ is size of the packet in bytes, $W$ be the radio bandwidth in bits per seconds and *T* be the time required by the adversary to successfully forge the MAC in number of days, then, *T* can be expressed as

$$T = \frac{2^{k-1} * P * 8}{W * 3600 * 24} \qquad (19)$$

Thus, this expression models the effort required by the adversary to carry out a cut-and-paste attack, based on the brute-force approach. We show the relation between T and the typical radio bandwidths in Fig. 19.

As can be observed from the graph, if we assume the radio is operating at 250 kbps, then for the same 68-bytes sized packets, an adversary needs 52 days (as compared to 621 days required with 19.2 kbps radio) of continuous packet transmission to forge the MAC. Therefore, the probability of a 4-byte MAC being forged by an adversary is higher than if higher MAC-sizes are used for the same. This is especially true in reference to the lifetime and the operating hours of the sensor nodes (left in the unattended environment) as shown in Fig. 20.

Thus, high data rate applications would demand more bytes to be allocated for MAC as they would naturally require higher bandwidth. An example of applications that demand so is Smart Kindergarten, wherein a WSN is deployed for monitoring the behavior of the children and their movements with video streaming [82].

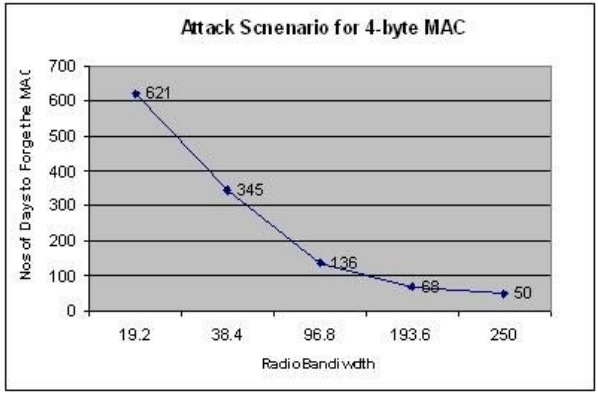

Figure 19 MAC-forging Attack Scenario with a 4-byte MAC relative to Radio Bandwidth

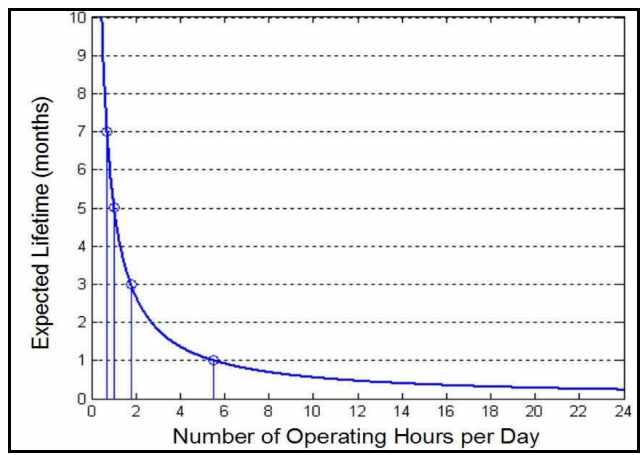

Figure 20 Lifetime of Typical Sensor Nodes [81]

In fact [10] gives a detailed dissection of the applications of WSNs in *object identification, detection & classification, single and multiple target tracking, collaborative signal & information processing* etc. that all - demand higher bandwidth.

In such applications, the volume of the data transferred as well as the frequency at which the data is transferred is high. Here, the probability of a 4-byte MAC being forged by an adversary operating with in the network is definitely higher than if an 8-byte MAC were employed.

Hence, we propose variable MAC sizes, which the application designer using our communication architecture, can choose, configure and implement in the application, deployed; thereby optimizing the level and type of security. Eventually, given a choice of the MAC sizes, the application designer using FlexiSec, can *choose, configure and implement* the application using appropriate MAC sizes, thereby optimizing the level and type of security.

Based on the above premises, we describe the modes of operations of FlexiSec briefly as below:

Mode 1: The first option in the architecture is named as *Null* wherein it is assumed that because the security support is implemented in the hardware, the proposed security framework in the operating system does not include any security feature. This option can be used while employing the security enabled radio chips for the WSNs.

Mode 2: The second one viz. *FlexiSecHASH* is proposed to offer the support only for message authentication – typically suited for the applications demanding only message integrity; without any demands for the entity authentication. Hence, we follow the un-keyed authentication technique i.e. hashing employing an algorithm like SHA-1. Any participating entity can check the authenticity of the message, irrespective of the keys employed or not.

Mode 3: *FlexiSecAUTH64* mode, the third mode, is intended to be employed for High Volume data and high data rate applications like PODS at Hawaii [83] OR Smart Sensors and Integrated Microsystems (SSIM) application [84] for Process control applications involving monitoring of machine parameters. Here, we intended to provide the support for 64 bits of MAC but without data encryption; because we believe confidentiality of data is not demanded here. This mode is suitable for the high data rate environmental monitoring applications, of the kind mentioned above.

Mode 4: Similarly, the fourth option viz. *FlexiSecAUTH32* offers *authentication-only* support for low data rate applications – with the provision of only 4 bytes of MAC. As mentioned earlier, the candidate applications are typically found in environmental control e.g. water-level monitoring, flood forecasting, stress monitoring in concrete structures etc.. In such applications, it is sufficient to sense and transmit only a few packets per day with only minimal parameter values.

Modes 5 & 6: The fifth and the sixth options viz. *FlexiSecAUTH_ENC64* and *FlexiSecAUTH_ENC32* are the options to support data confidentiality apart from the message integrity using 8-byte MAC and 4-byte MAC respectively. These modes are useful in the applications requiring confidentiality e.g. in the military applications of the kind referenced before.

Mode 7 & 8: The seventh and the eight modes in our proposed architecture are the *FlexiSecAUTH_REPP64* and *FlexiSecAUTH_REPP32* modes. These modes augment the security properties attained in the previous modes with that of replay protection. These modes could be employed for applications demanding message authentication as well as replay protection alone, without any encryption. Therefore, these modes are intended to employ CBC-MAC as the message authentication scheme.

The last mode of operation viz. *FlexiSec_AUTH_ENC_REPP64* basically is intended to offer all the security attributes listed above with a message TAG MAC of 8 bytes using the OCB mode again – could be employed in highly security-critical applications.

Thus, the proposed architecture indeed offers flexibly configurable security attributes for the applications.

Again, based on our performance results, we specify either XXTEA or AES cipher depending upon the availability of the resources. We also use the premise that if the application under consideration demands an 8-byte MAC, it must be implemented on the sensor node platform with higher resources. Therefore, we prescribe the AES cipher for the same. Otherwise the XXTEA cipher remains suitable.

## 8. Conclusion

The proposed flexible model of the link layer security architecture for the WSNs is useful in attaining the optimum performance in a deployed application. The resource optimization is possible because of the flexibility available in the hands of the application designer to select the specific security attributes as are demanded actually by the application. We believe that such a configurable link layer security architecture can always be useful in tuning the overhead associated with different WSN applications and thus make the entire system more responsive to the application environment. As compared to the peer link layer security architectures viz. TinySec, SenSec and MiniSec, in Table VI, we show the relative strengths of FlexiSec.

Thus, the overall contribution of this research work is in augmenting the link layer security framework for the WSNs with the new concept of configurability. We have focused here on configurability with respect to the applications and the available resources with a fixed Mica2 hardware platform. As the applicability of the WSNs is moving out of the research labs into the real-world domains, the research explorations carried out here need be implemented for different platforms, too.

In addition, we have used *uniform global keying* for key deployment and management. However, in order to support scalability in the real world applications, appropriate keying mechanism using the sophisticated keying mechanism based on either the SKC or PKC need to be devised as are described in [85].

We believe that the time is ripe for incorporating sound security mechanisms in their implementations. Instead of discovering the limitations in the post-deployment implementations, it is better to impose the improvements in the designs and test them. We believe that the research findings attempted here will go a long way in achieving this objective.

## Acknowledgment

We thank all those anonymous reviewers who have devoted significant time and efforts to bring this paper to the shape it is in, now.

**Devesh Jinwala** was born on 3$^{rd}$ July 1964. He has a Master's degree in Electrical Engineering from the Maharaja Sayajirao University of Baroda, India with specialization in Microprocessor Systems and Applications.

He is employed as an Assistant Professor in Computer Engineering with Sardar Vallabhbhai National Institute of Technology, Surat (India) since 1991. He is currently working on Configurable Link layer Security Protocols for Wireless Sensor Networks. His major areas of interest are Information Security Issues in Resource Constrained Environment, Algorithms & Computational Complexity and Software Engineering.

**Dhiren Patel** was born on 29$^{th}$ July 1966. He has a Master's degree in Computer Science & Engineering from IIT Kanpur, India and Ph D in Computer Engineering from the South Gujarat University (REC Surat), India.

He is employed as a Professor of Computer Engineering at NIT Surat, India. His major areas of interest are Information/Network Security, Web Engineering and Ubiquitous Architectures.  Apart from his numerous International publications and distinguished talks, He has also authored a book "Information Security: Theory & Practice" published by Prentice Hall of India in 2008

**Kankar Dasgupta** was born on 14$^{th}$ September 1951. He has a Masters Degree in Computer Science and Engineering from Jadavpur University, Kolkata and Doctorate in Electrical Engineering from the Indian Institute of Technology, Bombay.

He is currently the Director of DECU at the Indian Space Research Organization, Ahmedabad. His research areas are Satellite Imaging, Digital Communication and Networks. He has been associated with many academic institutes in guiding them deliver quality research outputs.

Table V    FlexiSec modes of operations

| Sr No | Mode Identifier | Description | Cipher to be Employed | |
|---|---|---|---|---|
| | | | Low Storage/Energy | Higher Storage/Energy |
| 1. | Null | Security support in hardware radio chip | - | - |
| 2. | FlexiSecHASH | Naïve Authentication Support with one-way hash function SHA-1 | XXTEA cipher | AES cipher |
| 3. | FlexiSecAUTH64 | 64 bits - 8 bytes – MAC : only keyed authentication – CBCMAC | XXTEA cipher | AES cipher |
| 4. | FlexiSecAUTH32 | 32 bits - 4 bytes – MAC : only keyed authentication – CBCMAC | XXTEA cipher | AES cipher |
| 5. | FlexiSecAUTH_ENC64 | 8 bytes MAC and encryption – OCB (single pass) | XXTEA cipher | AES cipher |
| 6. | FlexiSecAUTH_ENC32 | 4 bytes MAC and encryption – OCB (single pass) | XXTEA cipher | AES cipher |
| 7. | FlexiSecAUTH_REPP64 | 8 bytes MAC: keyed authentication (CBCMAC) & replay protection | XXTEA cipher | AES cipher |
| 8. | FlexiSecAUTH_REPP32 | 4 bytes MAC: keyed authentication (CBCMAC) & replay protection | XXTEA cipher | AES cipher |
| 9. | FlexiSec_ AUTH_ENC_REPP64 | 8 bytes MAC: keyed authentication (OCB), encryption & replay protection | XXTEA cipher | AES cipher |

Table VI   FlexiSec Link Layer Security Framework vs. TinySec, SenSec, MiniSec Security Link Layer Frameworks

| Sr No | Support for Link Layer Security Framework Features viz. | | TinySec | SenSec | MiniSec | FlexiSec |
|---|---|---|---|---|---|---|
| 1. | Confidentiality? | | Yes | Yes | Yes | Yes |
| 2. | Message/Entity Authentication? | | Yes | No | No | Yes |
| 3. | Authenticated Encryption? | | Conventional | Conventional | True | True |
| 4. | Variable MAC-sizes | | No | No | No | Yes |
| 5. | Modular Design? | | Yes | Yes | Yes | Yes |
| 6. | Replay Protection Algorithm? | | No | No | Yes | Yes |
|  | (a) | Counter-based ? | - | - | Yes | Yes |
|  | (b) | SHA-1 based ? | - | - | No | Yes |
|  | (c) | Bloom-filter based ? | - | - | Yes | Yes |
| 7. | Configurable? | | No | No | No | Yes |
| 8. | Hardware Platform? | | Mica2 | Mica2 | Telos | Mica2 |
| 9. | Underlying OS? | | TinyOS | TinyOS | TinyOS | TinyOS |